\documentclass[sigconf]{acmart}

\usepackage{amsmath,amsfonts}
\usepackage{algorithmic}
\usepackage{graphicx}
\usepackage{textcomp}
\usepackage{xcolor}
\usepackage{tcolorbox}
\usepackage{graphicx}
\usepackage{listings}
\usepackage{makecell}
\usepackage{adjustbox}
\usepackage{multirow}
\usepackage{graphicx}
\usepackage{svg}
\usepackage{xspace}
\usepackage{amsmath}
\usepackage{subfigure}
\usepackage{caption}
\usepackage{amsmath}
\usepackage{booktabs}
\usepackage{color, colortbl}
\usepackage{url}
\usepackage{caption} 

\DeclareUnicodeCharacter{2212}{-}
\definecolor{Gray}{gray}{0.9}
\definecolor{green}{RGB}{102,252,102}
\definecolor{ored}{RGB}{255,99,71}
\definecolor{orange}{RGB}{255,165,0}
\definecolor{lightgray}{RGB}{211,211,211}

\usepackage[ruled,vlined,linesnumbered]{algorithm2e}

\SetCommentSty{mycommfont}

\usepackage{tikz}
\usetikzlibrary{shapes}

\newcommand{\tool}{\texttt{Gallery D.C.}\xspace}

\AtBeginDocument{%
  \providecommand\BibTeX{{%
    \normalfont B\kern-0.5em{\scshape i\kern-0.25em b}\kern-0.8em\TeX}}}

\copyrightyear{2022}
\acmYear{2022}
\setcopyright{acmcopyright}\acmConference[ICSE '22 Companion]{44th International Conference on Software Engineering Companion}{May 21--29, 2022}{Pittsburgh, PA, USA}
\acmBooktitle{44th International Conference on Software Engineering Companion (ICSE '22 Companion), May 21--29, 2022, Pittsburgh, PA, USA}
\acmPrice{15.00}
\acmDOI{10.1145/3510454.3516873}
\acmISBN{978-1-4503-9223-5/22/05}

\begin{document}

\title{Gallery D.C.: Auto-created GUI Component Gallery for Design Search and Knowledge Discovery}

\author{Sidong Feng}
\affiliation{%
  \institution{Monash University}
  \city{Melbourne}
  \country{Australia}}
\email{sidong.feng@monash.edu}

\author{Chunyang Chen}
\affiliation{%
  \institution{Monash University}
  \city{Melbourne}
  \country{Australia}}
\email{chunyang.chen@monash.edu}

\author{Zhenchang Xing}
\affiliation{%
  \institution{Australian National University}
  \city{Canberra}
  \country{Australia}}
\email{zhenchang.xing@anu.edu.au}

\renewcommand{\shortauthors}{Feng, et al.}

\begin{abstract}
GUI design is an integral part of software development. 
The process of designing a mobile application typically starts with the ideation and inspiration search from existing designs.
However, existing information-retrieval based, and database-query based methods cannot efficiently gain inspirations in three requirements: design practicality, design granularity and design knowledge discovery.
In this paper we propose a web application, called \tool that aims to facilitate the process of user interface design through real world GUI component search.
\tool indexes GUI component designs using reverse engineering and deep learning based computer vision techniques on millions of real world applications.
To perform an advanced design search and knowledge discovery, our approach extracts information about size, color, component type, and text information to help designers explore multi-faceted design space and distill higher-order of design knowledge.
\tool is well received via an informal evaluation with 7 professional designers.

Web Link: \url{http://mui-collection.herokuapp.com/}.

Demo Video Link: \url{https://youtu.be/zVmsz_wY5OQ}.
\end{abstract}



\maketitle

\section{Introduction}
Graphical User Interface (GUI) is ubiquitous in almost all modern desktop software, mobile applications and online websites. 
It provides a visual bridge between a software application and end users through which they can interact with each other. 
A good GUI design makes an application easy, practical and efficient to use, which significantly affects the success of the application and the loyalty of its users~\cite{jansen1998graphical,feng2021auto,chen2021should,zhao2021guigan}. 
To design a good GUI, designers usually starts with the ideation and inspiration search based on the initial user needs and software requirements~\cite{chen2020lost}.

Despite the enormous amount of GUI design search engines existed, it is still difficult for designers to efficiently get inspiration and understand design knowledge.
Three requirements behind this phenomenon: \textit{design practicality}, \textit{design granularity} and \textit{design knowledge discovery}. 
First, designers want to see the practical use of GUI designs in real applications, not just artworks, as GUI designs in artwork and real application have substantial differences.
Second, designers want to see both the overall design and the detailed design of the GUI components. Although there are some GUI component design kits, but they do not provide the GUI context in which certain components can be practically composed. 
Third, designers want advanced GUI design search capabilities and knowledge discovery support.

To address this, we design and implement \tool \footnote{\tool is an abbreviation for \textbf{Gallery} of \textbf{D}esign \textbf{C}omponents}, a gallery of 11 types of GUI design components (e.g., button, image button, check box, radio button, switch, toggle button, progress bar, rating bar, seek bar, spinner, and chronometer) that harness GUI designs crawled from millions of real-world applications using deep learning based reverse-engineering and computer vision techniques. 
\tool assists designers and developers to (1) access 68k practical GUI designs, 469k app-introduction GUI designs and 32k GUI component designs of 130k Android apps (2) enable multi-faceted search (e.g., component types, size, color, application category and/or displayed text) on 11 types component design (3) understand advanced design knowledge discovery through visualising design demographics ( distributions of component type usage, primary colors, and height/width) and design comparison features.

\begin{figure*}
  \centering
  \includegraphics[width=0.95
  \textwidth]{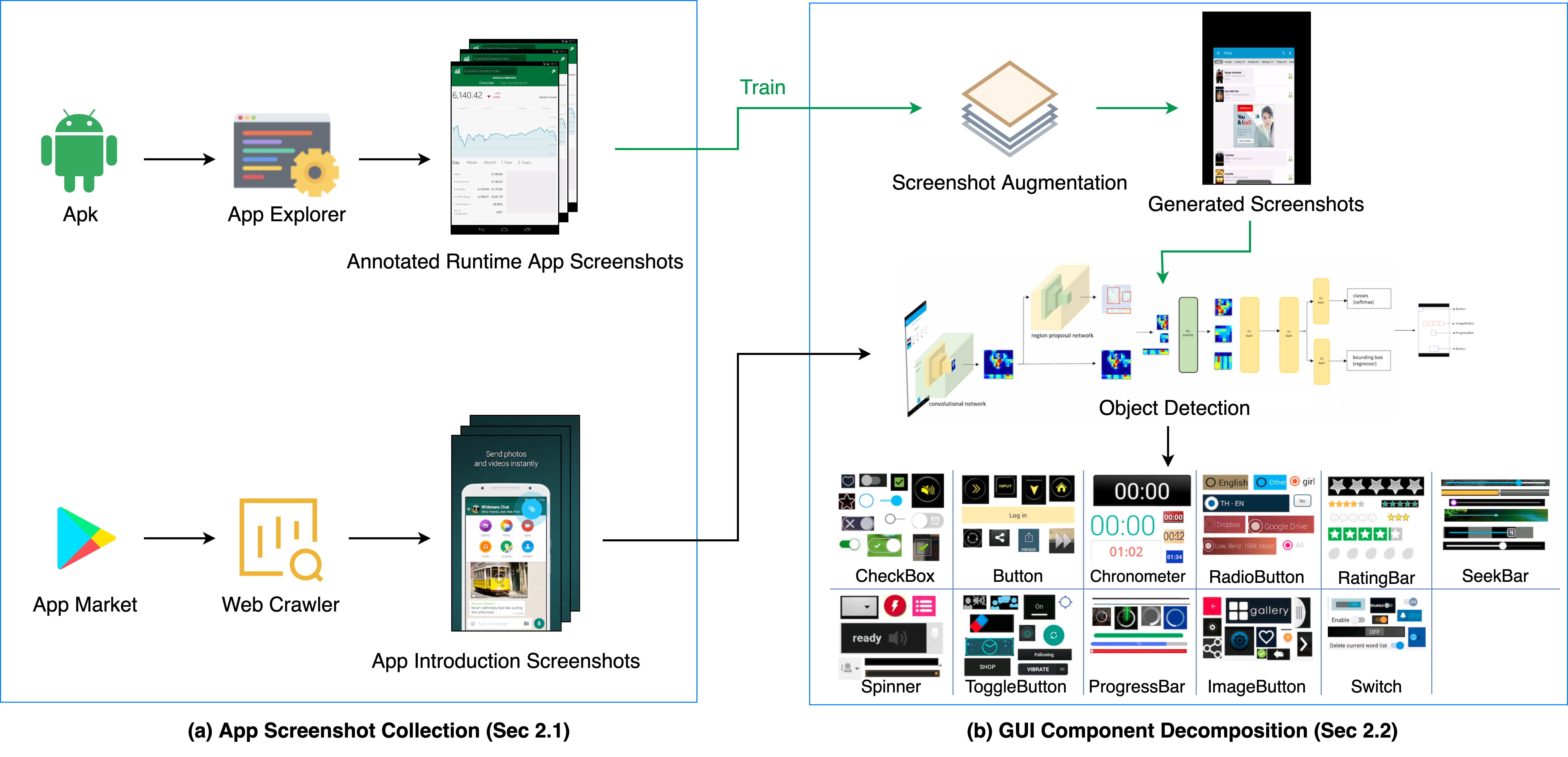}
  \caption{The overview of GUI component dataset collection for \tool.}
  \label{fig:approach}
\end{figure*}

\tool represents a significant departure from and improvement over existing works and websites. 
For example, Dribbble~\cite{web:dribble} only shares GUI design artworks to exchange creativity and aesthetics.
In contrast, our work exploits real world design resources, e.g., screenshots of real apps and app introduction in the application market.
There are many related works on GUI design search~\cite{chen2020wireframe,liu2020owl,chen2020unblind,chen2018ui} but rarely related to GUI component search.
The most closely related approach to our tool is Guigle~\cite{bernal2019guigle} that utilizes automated information retrieval techniques and indexes metadata to return relevant GUI designs in relation to a query.
While the Guigle approach represents a promising technique for helping developers to search GUI designs in granularity, it does little to help designers get inspiration on components as it returns GUI designs at the whole GUI level.
To the best of our knowledge, our work is the first to build a large-scale GUI \textbf{component} design gallery using app introduction GUI screenshots.

This paper makes the following contributions:
\begin{itemize}
    \item {A new gold mine of design resource, e.g., app introduction GUI screenshots usually illustrate the most important features and the best-designed GUIs of an application.}
    \item{We provide an opportunity for designers to learn about GUI designs, gain design inspiration and understand design trend on a massive scale. }
\end{itemize}

\section{Dataset Collection}
Figure~\ref{fig:approach} provides an overview of dataset collection for \tool, including two main steps: app GUI screenshot collection and GUI component wirification.
More details can be seen in our previous work~\cite{chen2019gallery}.

\subsection{App GUI Screenshot Collection}
We collected two kinds of app GUI screenshots, annotated runtime app screenshot and app introduction screenshot. 

\textbf{Annotated Runtime App Screenshot}:
We downloaded Android apps (e.g., APKs) from Google Play and dumped app metadata including app category, download times, etc. Based on the Android application GUI test framework, we developed an automated GUI exploration method which can automatically execute a mobile application and explore the GUIs of app by simulating user interaction. During such automated GUI explorations, we used Android UI Automator~\cite{web:UIautomator} to dump runtime GUI screenshots and its corresponding GUI component metadata (e.g., component type, size and position).
We further leveraged Rico dataset~\cite{deka2017rico} to extend our dataset.
This dataset contains 68k screenshots from 4k apps.

\textbf{App Introduction Screenshot}:
The app introduction screenshot illustrates the most important GUIs of an app (usually the one with the best design), therefore, we were motivated to exploit this undiscovered GUIs gold mine to extend our gallery.
We built a web crawler based on the breadth-first search strategy~\cite{najork2001breadth} e.g., collecting a queue of URLs from a seed list, and putting the queue as the seed in the second stage while iterating. Apart from the app introduction designs, our crawler also dumped its corresponding app metadata. This dataset contains 469k app introduction GUI screenshots of 126k Android apps. 

\subsection{GUI Component Decomposition}
It is straightforward to decompose components from annotated runtime app screenshot which contains components metadata, however, the app introduction screenshots are purely images without any hierarchy information. Therefore, we proposed an approach to decompose GUI components from app introduction screenshots.

\begin{figure*}[ht!]
  \centering
  \includegraphics[width=0.95\textwidth]{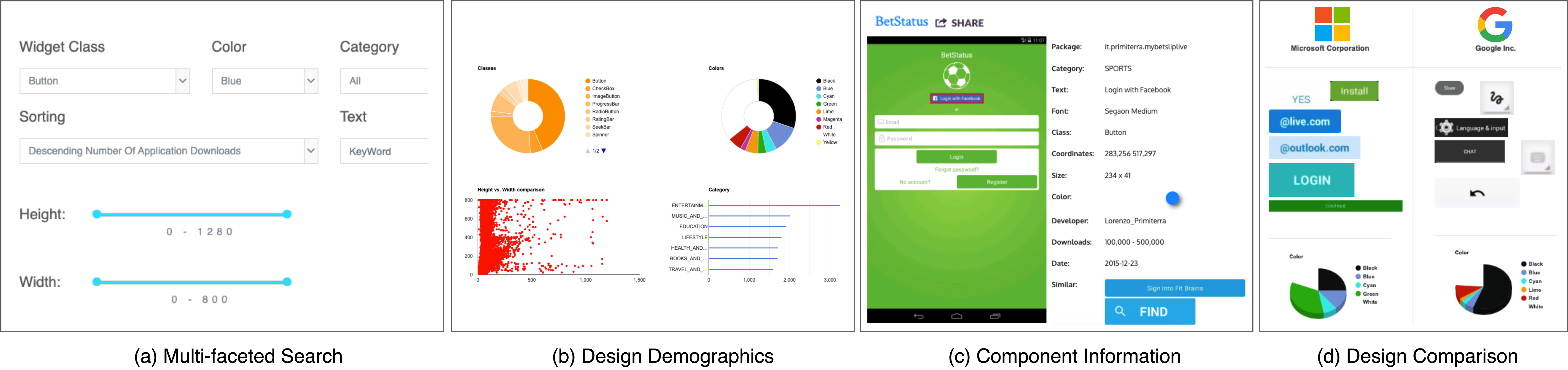}
  \caption{Illustration of our \tool web application.}
  \label{fig:web}
\end{figure*}

\textbf{Screenshot Augmentation}:
The app introduction screenshots are more like posters which contain not only the app GUI, but also many other information such as app name, short descriptions.
To reduce such difference, we first applied a GUI-screenshot specific image augmentation method (randomly adjust the size and position) to simulate the annotated runtime screenshots as the app introduction screenshots.

\textbf{Object Detection}:
We then used the simulated annotated runtime screenshots to train a GUI component detection model using Faster RCNN~\cite{DBLP:journals/corr/RenHG015}. 
The model used convolutional neural network (CNN) to extract image features and region proposal network (RPN) to generate Region of Interest (ROI). At the end, the model applied an object detection model to detect and extract different components. 
Thus, our trained model can automatically decompose the app introduction screenshots into a set of GUI components and determine the component type, position, width and height.

\section{TOOL IMPLEMENTATION AND USAGE}

\tool is a web app, which provides several convenient techniques for users to search, analyse, compare the GUI designs to gain inspirations.

\textbf{Multi-faceted Search (Figure~\ref{fig:web}a)}:
We allow the users to search for component class,  category, height, and width, and these attributes can be easily indexed by metadata.
Color design is an important aspect of GUI design.
In order to perform search for color, we augmented primary color of GUI component by using HSV colorspace detection. 
Moreover, text search query allows the users to find all components of a text and visualize their design styles.
To achieve this, we used Optical Character Recognition (OCR) technique to extract the text information on the GUI component.

\textbf{Design Demographics (Figure~\ref{fig:web}b)}:
To help designers understand the overall design landscape at an aggregation level, we dynamically compute design demographics for the multi-faceted search results, so that the designers can obtain a quick overview of the key characteristics of the search results. 
The system computes four types of design demographics: the component color (computed by HSV colorspace), the component height and width (obtained from object detection), the number of components of each component type (determined by object detection), the number of component of each application category (obtained from metadata).
The system adopts Google Analytics Chart api to visualize the distribution in piechart (component class and color), scatter plot (component height and width), and bar chart (app category).

\textbf{Component Information (Figure~\ref{fig:web}c)}:
Once the user finds an attractive component, he can click on it to view more detailed information, such as the whole GUI (component is highlighted with red bounding box), app package, developers, etc.
We provide users with some similar components to extend their design inspirations.
The selection of similar components is calculated by the correlation of app, developer, color, text, etc.
We also support the users to put the component into their design notebook for future reference, using "Share" button.

\textbf{Design Comparison (Figure~\ref{fig:web}d)}:
Design comparison allows designers to see the brand commonalities and variations of different companies.
We borrowed the idea of comparison shopping to support design comparison. 
The gallery system allows designer to select different companies to compare their GUI component designs. 
Note that only high reputation/ranking companies are displayed (34 companies). As a company has many apps, and an app has many components, we only displayed the top six components from app with high downloads and high reputation.
If the company does not have a particular type of components, the system reports "None" for this component type. The system also computes the color distribution of components from the company.
Finally, it generates a comparison table for designers to visually compare GUI components side by side.

\textbf{User Experience (UX)}:
To build a scalable website, we adopted Node.js (Express as the framework) as it is asynchronous and event-driven. 
Since we had a large number of database, the efficiency of our website is crucial.
We applied two methods. 
First, for the database side, we adopted MongoDB as our database program, because on the one hand, it is well integrated with Node.js; on the other hand, it enables more efficient large-scale processing of data.
Second, we used the lazy loading strategy which is commonly used in image search engines. It showed some initial search results on the visible screen and loaded more search results when users scroll down the search results.
The core technique of this strategy is an event listener which listens to the events in the browser such as scrolling, resizing, etc. After we recognized a change, we determined the space in the browser and calculated the number of images that should become visible on the screen using window height, image's top offset, etc. Finally, we triggered them to load accordingly.

To keep the style of web pages consistent, we adopted the most popular CSS framework, Bootstrap.
For the search page, to optimize the use of space when displaying components, we adopted a widely used layout, masonry layout.
Instead of the common vertical masonry, we decided to set masonry in horizontal as most of the components are flat and wide. 
\section{Evaluation}
The goal of this section is to evaluate our platform \tool in terms of (i) its performance in decomposing GUI components, and (ii) the usefulness of our \tool in a real design environment.
We elaborate more details and illustrations in our previous work~\cite{chen2019gallery}.


\subsection{Performance of Decomposition}
Among all 68,702 annotated runtime GUI screenshots, we leveraged 90\% to train our object detection model.
We empirically set the training configurations following~\cite{chen2020object,xie2020uied}.
To evaluate the performance of the model, we employed three principal metrics, i.e., precision, recall, mean Average Precision (mAP).
We set IoU to 0.8 as the metric threshold. 

We first evaluated our model on the remaining 10\% testing dataset of annotated runtime GUI screenshots.
The overall performance is 0.62 for recall, 0.76 for precision, and 0.51 for mAP.
We further checked the performance of our model on the app introduction screenshots. 
As there is no groundtruth for app introduction screenshots, we manually annotated GUI components in 100 screenshots.
Our model achieves 0.53, 0.84, 0.62 for recall, precision and mAP, respectively.
In addition, adding screenshot augmentation can boost the performance of decomposing GUI component in app introduction screenshot by 43.2\% for recall, 5\% for precision, and 44.2\% for mAP.

\subsection{Usefulness of \tool}
We surveyed 7 professional designers\footnote{2 Google, 1 Facebook, 1 Alibaba, 1 Huawei, 1 Volkswagen-Mobvoi, 1 TAL education} to assess the usefulness of \tool. 
The designers responded positively to the use of website.
We collected and analysed their general feedback on how our website can be used for their own application design tasks.

\textbf{Usage Scenario 1: Design Granularity.}
One designer works on designing an application for game casual. 
One of her major job is to decide the \textit{button} style of the game application. She praised that our gallery is particularly useful for providing inspirations to her. Since the main target audience for her app is young girls, she was looking for buttons that fit young girl’s style. One button of pink color with reflective bubble was selected by her as this kind of design is deemed to attract young girls. To remind her of this design inspiration, she put the design in her design notebook for reference. She also found another button with ranking-list icon interesting. It inspired her not only the button semantic, but also reminded her to add a ranking mechanism into the game. In addition to the button design itself, she also learned where the button is composed in the real-world app and embeds into her own app. 

\tool explores a new search strategy for gaining inspirations, bottom-up search. Search from components to the whole, and then gradually determine and expand design styles.

\textbf{Usage Scenario 2: Design Practicality}
One designer is creating a GUI component design system, which is a set of standards for designing and coding with components. This GUI component design system can enhance the visual and interactive consistency of different products within the company to provide a better and familiar user experience. 
In order to understand the design style of different companies, he adopted the comparison function provided in our \tool. He looked at two large companies (\textit{Google} and \textit{Microsoft}) for a detailed comparison.
The comparison table presented representative GUI components of each company side by side.
By this intuitive visual comparison, he clearly discovered that \textit{Microsoft} prefers using right-angle rectangle components, while \textit{Google} prefers using white, black and gray color as the dominant color and furthermore embedding the shadowing effects to the components. 
With those features displayed, he further thought about how to distinguish his design system from the current ones.

\tool does help designers to understand the design variations between companies via practical GUI components and facilitate their own designs.

\textbf{Usage Scenario 3: Design Knowledge Discovery}
One designer also provided another example of using our website to help the design of a social application. 
She searched the GUI components by \textit{social} application category in Gallery D.C. 
On the one hand, lots of component designs were retrieved for her to gain inspirations. On the other hand, four demographics were reported to help her understand design knowledge. She observed that most \textit{buttons} within the social-media apps are rather flat and wide as compared with other categories.
After reviewing her experience in using social apps and searching the web, she further verified that this is a common phenomenon in social apps, because wide and large buttons can easily attract the attention of the audience, especially for the widespread young users in social apps.

\tool equips with the summarization of GUI component designs is a powerful approach to understand the design trend and practices.
\section{conclusion}
In this paper, we present \tool, a novel GUI component design gallery system that can satisfy the designers' information need for \textit{design practicality}, \textit{design granularity}, and \textit{advanced design search and knowledge discovery}. We exploit a gold mine of app introduction screenshots in the app market to further extend our component gallery, using deep learning model and computer vision techniques. Finally, \tool receives positive feedback from 7 professional designers who see its potential as an useful tool for their real design projects.

In the future, we will further remove some low-quality data within our gallery as the designers found that some retrieving results do not look professional, hence no reference value. Furthermore, several designers wonder whether our gallery could extend to other platforms, such as IOS, web.

\bibliographystyle{ACM-Reference-Format}
\bibliography{main}
\end{document}